\documentclass[preprint,12pt,authoryear]{elsarticle}

\usepackage{amssymb}
\usepackage{amsthm}
\usepackage{listings}
\usepackage[svgnames]{xcolor}
\usepackage{subcaption}
\usepackage{caption}

\usepackage{amssymb}
\usepackage{amsthm}
\usepackage{amsmath}
\usepackage{bm}
\usepackage{booktabs}
\usepackage{hyperref}
\usepackage{natbib}
\usepackage{lineno}
\usepackage{verbatim}
\usepackage{pgf, tikz}
\usetikzlibrary{arrows,automata,fit}
\newcommand{\indep}{\perp \!\!\! \perp}
\usepgflibrary{shapes.geometric}
\newcommand{\stages}[2]{\tikz{\node[shape=circle,draw,inner sep=1pt,fill=#1,minimum size=0.5cm]{${#2}$};}} 
\newcommand{\xx}{1}
\newcommand{\yy}{1}
\usepackage{graphicx}

\journal{}

\begin{document}

\begin{frontmatter}



\title{Sensitivity and robustness analysis in Bayesian networks with the bnmonitor R package}


\author[ie]{Manuele Leonelli}
\author[bo]{Ramsiya Ramanathan}
\author[rlw]{Rachel L. Wilkerson}

\address[ie]{School of Human Sciences and Technology, IE University, Madrid, Spain.}
\address[bo]{Dipartimento di Scienze Statistiche, Universit\`{a} di Bologna, Bologna, Italy. }
\address[rlw]{Tesserwell, LLC}

\begin{abstract}
Bayesian networks are a class of models that are widely used for risk assessment of complex operational systems. There are now multiple approaches, as well as implemented software, that guide their construction via data learning or expert elicitation. However, a constructed Bayesian network needs to be validated before it can be used for practical risk assessment. Here, we illustrate the usage of the \texttt{bnmonitor} R package: the first comprehensive software for the validation of a Bayesian network. An applied data analysis using  \texttt{bnmonitor} is carried out over a medical dataset to illustrate the use of its wide array of functions.
\end{abstract}


\end{frontmatter}


\section{Introduction}
\label{sec:introduction}

Assessing the validity of a statistical model is a fundamental step of any real-world applied analysis to ensure that the conclusions drawn from the model are reliable and credible \citep{Razavi2021,Saltelli2000}. Depending on the research community or the model used, this analysis step may take different names such as model validation, model checking, sensitivity analysis or robustness checking.

Here we focus on the model class of Bayesian networks (BNs) \citep{Darwiche2009,Koller2009}, the most commonly used probabilistic graphical model, which gives an intuitive visualization of the dependence structure between variables as interest as well as an efficient platform to answer inferential queries. The array of domains where BNs are used is constantly increasing \citep[e.g.][]{Akhavan2021,Bielza2014,Cai2018,Chen2021,Drury2017,Mclachlan2020}.

Although methods for model validation in BNs have been developed \citep[][among others]{Chan2002,Coupe2002,Cowell2007,Pitchforth2013,Rohmer2020}, their use in practice has been limited, although it has been increasing in the past few years \citep[e.g.][]{Chen2012good,Hanninen2012,Kleemann2017,Makaba2021}.
Conversely, for other modelling approaches model validation is basically always carried out. For instance, in linear regression modelling one always checks the distribution of the residuals to assess if the model's assumptions are met. In Bayesian inference, posterior predictive checks compare the original dataset to one simulated from the predictive model distribution to assess if the model is appropriate.

A possible reason behind the limited use of model validation techniques in BNs may be the lack of implemented methods in many software \citep[see][for an exception]{samIam}. We have recently developed the  \texttt{bnmonitor} R package to fill this gap and provide practitioners with a wide array of functions to check the validity of their model. We chose the R programming language for two reasons. First, because there are now a large number of packages to both learn BNs from data and to carry out inferential tasks, including \texttt{bnlearn} \citep{Scutari2010}, \texttt{BayesNetBP} \citep{Yu2020} and \texttt{gRain} \citep{Hojsgaard2012}. Second, because we can take advantage of the advanced graphical capabilities of R, by using methods from the \texttt{ggplot2} \citep{Wickham2016} and \texttt{qgraph} \citep{qgraph} packages.

Here we provide an overview of the capabilities of \texttt{bnmonitor} by carrying out an applied BN sensitivity analysis over a medical dataset. Although \texttt{bnmonitor} requires as input BNs as \texttt{bnlearn} objects, the latter package has a wide array of conversion functions from formats used in most other software. Therefore, \texttt{bnmonitor} can be used in conjunction with models developed in any other programming language or commercial software.  Before carrying out the analysis, we give a brief introduction to both BNs and the model validation techniques implemented in \texttt{bnmonitor}. Although \texttt{bnmonitor} can consider BNs with either discrete variables or continuous ones under the Gaussian assumption, for the purpose of this paper we consider only the discrete case, since this is the most common in practical applications. We refer to \citet{Gorgen2020} for details on the continuous case.

\section{Model validation in Bayesian networks}\label{sec:bn}

Let $[n]=\{1,2,\dots,n\}$. Consider a random vector $Y =(Y_1,\dots,Y_n)$ of interest, where $Y_i$ takes values in a discrete space $\mathbb{Y}_i$, $i\in[n]$. For a subset $A \subset [n]$, we denote $Y_{A}=(Y_i)_{i\in A}$ and $\mathbb{Y}_{A}=\times_{i\in A}\mathbb{Y}_i$. For three random vectors $Y_{A}$, $Y_{B}$ and $Y_{C}$, where $A, B, C \subset [n]$, we say that $Y_{A}$ is conditionally independent of $Y_{B}$ given $Y_{C}$ and write $Y_{A}\indep Y_{B}|Y_{C}$ if 
\[
p(Y_A=y_A|Y_{B}=y_{B},Y_{C}=y_{C})=p(Y_{A}=y_{A}|Y_{C}=y_{C}),
\]
for all $y_{A}\in\mathbb{Y}_{A}$, $y_{B}\in\mathbb{Y}_{B}$ and $y_{C}\in\mathbb{Y}_{C}$, where $p$ denotes a probability mass function. In the following, as a shorthand, we write $p(Y_{A}=y_{A}|Y_{B}=y_{B})$ as $p(y_{A}|y_{B})$, for any sets $A, B \subseteq[n]$.

\subsection{Bayesian networks}
A BN gives a visual representation of conditional independence by means of a directed acyclic graph (DAG). The DAG associated to the BN has vertex set $[n]$, i.e. a vertex is associated to each variable of $Y$, and edges represent dependence. More formally, a BN for a discrete random vector $Y$ consists of:
\begin{itemize}
\item $n-1$ conditional independences of the form $Y_i\indep Y_{[i-1]}|Y_{\Pi_i}$ where $\Pi_i\subset [i-1]$;
\item a DAG $G$ with vertex set $[n]$ and edge set $\{(j,i): i\in[n], j\in \Pi_i\}$;
\item conditional probabilities $p(y_i|y_{\Pi_i})$, for $y_i\in\mathbb{Y}_i$ and $y_{\Pi_i}\in\mathbb{Y}_{\Pi_i}$.
\end{itemize}

The vector $Y_{\Pi_i}$ includes the \emph{parents} of the variable $Y_i$, i.e. those variables $Y_j$ such that there is an edge $(j,i)$ in the DAG $G$ of the BN. Given these conditional probabilities, the probability distribution $p_G$ of the BN can be derived as
\[
p_G(y) = \prod_{i=2}^np(y_i|y_{{\pi}_i})p(y_1)
\]
\begin{figure}
\centering
\begin{tikzpicture}
\renewcommand{\xx}{2}
\renewcommand{\yy}{1.5}
\node (1) at (0*\xx,0*\yy){\stages{white}{1}};
\node (2) at (0*\xx,1*\yy){\stages{white}{2}};
\node (3) at (1*\xx,0*\yy){\stages{white}{3}};
\node (4) at (1*\xx,1*\yy){\stages{white}{4}};
\node (5) at (2*\xx,0.5*\yy){\stages{white}{5}};
\draw[->, line width = 1.1pt] (1) -- (3);
\draw[->, line width = 1.1pt] (2) -- (4);
\draw[->, line width = 1.1pt] (3) -- (5);
\draw[->, line width = 1.1pt] (4) -- (5);
\draw[->, line width = 1.1pt] (1) -- (4);
\end{tikzpicture}
\caption{A simple DAG of a BN with vertex set $\{1,2,3,4,5\}$ and edge set $\{(1,3),(1,4),(2,4),(3,5),(4,5)\}$.\label{fig:bn}}
\end{figure}
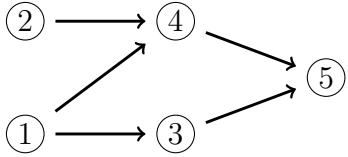

Figure \ref{fig:bn} reports a DAG over five discrete variables $(Y_1,\dots,Y_5)$. This DAG embeds the conditional independences $Y_2\indep Y_1$, $Y_3\indep Y_2|Y_1$, $Y_4\indep Y_3| Y_1,Y_2$ and $Y_5\indep Y_1,Y_2| Y_3,Y_4$. The definition of the BN is completed by a numerical specification of the probabilities $p$ of each variable conditional on the possible values of the parents. Its factorization can be written as
\[
p_G(y)=p(y_5|y_4,y_3)p(y_4|y_2,y_1)p(y_3|y_1)p(y_2)p(y_1)
\]

The DAG associated to a BN provides an intuitive overview of the relationships existing
between variables of interest. However, it does also provide a framework to assess if any
generic conditional independence holds for a specific subset of the variables via the so-called
d-separation criterion \citep[see e.g.][]{Darwiche2009}. Furthermore, the DAG provides a
framework for fast exact propagation of probabilities and evidence for the computation of
any (conditional) probability involving a specific subset of the variables.

\subsection{Robustness to data in Bayesian networks}
\label{sec:rob}
Often in practice a BN is learned from data. By learned, we mean that both its DAG $G$ and the associated probabilities $p$ are derived from a dataset. For the task of learning $G$,  novel algorithms continue to appear to account for more flexible structures and to improve speed and accuracy \citep[e.g.][]{Dai2020,Wang2021}. For learning $p$ given a DAG $G$, either a frequentist or a Bayesian approach can be taken. The \texttt{bnmonitor} package provides functions that quantify how well a learned BN actually represents the data and for which variables the model fit is poorer.

Suppose a dataset $\pmb{y} = (\pmb{y}_1,\dots,\pmb{y}_m)$ of $m$ observations of the random vector $Y$ has been collected, where $\pmb{y}_i=(y_{i1},\dots,y_{in})$ and $y_{ij}$ denotes the $i$-th observation for the $j$-th variable. We further let $\pmb{y}_{[i-1]}=(\pmb{y}_1,\dots,\pmb{y}_{i-1})$ and $\pmb{y}_{i}^{-j}=(y_{i1},\dots,y_{i(j-1)},y_{i(j+1)},\dots,y_{ip})$.

The following \emph{monitors} are used to assess the model fit of a BN to data. 

\paragraph{Global monitor}
The global monitor is the contribution of each vertex to the negative log-likelihood of the BN. Formally, the global monitor for $Y_j$ and a BN $G$ is 
\[
-\sum_{i=1}^m\log(p(y_{ij}|\pi_{ij}))
\]
where $\pi_{ij}$ is the value of the parents of $Y_j$ for the $i$-th observation. Higher values of the global monitor indicate vertices that had a bigger impact in the DAG selection process.

\paragraph{Sequential node monitors}
These diagnostics fall within the famous prequential framework of \citet{Dawid1992} and were further developed in \citet{Cowell2006,Cowell2007}. The node monitor assesses the adequacy of the marginal and
conditional probability distributions for each node in the model.  Let $p_i$ denote the predictive density of the BN learned using the dataset $\pmb{y}_{[i-1]}$ including only the first $i-1$-th observations (\ref{sec:app} gives details on how to compute predictive densities). The level of surprise of observing the value $y_{ij}\in\mathbb{Y}_j$ for the $j$-th variable after having processed $i-1$ observations, i.e. $\pmb{y}_{[i-1]}$, is given by the logarithmic score
\begin{equation}
\label{eq:score}
   S_{ij} = -\log(p_i(y_{ij})).
\end{equation}

Next, the logarithmic score is normalized. Define
\begin{equation}
\label{eq:exp}
    E_{ij} = -\sum_{y_j\in\mathbb{Y}_j}p_i(y_j)\log(p_i(y_j)), \hspace{1cm} V_{ij}=\sum_{y_j\in\mathbb{Y}_j}p_i(y_j)\log(p_i(y_j))^2-E_{ij}^2,
\end{equation}
where $E_{ij}$ and $V_{ij}$ are the expectation and the variance of the logarithmic score in Equation (\ref{eq:score}). The sequential marginal node monitor for $Y_j$ after having processed $\pmb{y}_{[i-1]}$ is defined as 
\begin{equation}
\label{eq:mon}
    Z_{ij}=\frac{\sum_{k=1}^iS_{kj}-\sum_{k=1}^iE_{kj}}{\sqrt{\sum_{k=1}^iV_{kj}}}
\end{equation}
For sufficiently large sample sizes, under the model assumptions, $Z_{ij}$ follows a standard Normal distribution if the model could have plausibly generated the data. Therefore, for instance, values of $Z_{ij}$ over 1.96 in absolute value may be an indication of poor model fit.

Another node monitor is similarly defined and called sequential conditional node monitor. This is defined as in Equations (\ref{eq:exp})-(\ref{eq:mon}), but the logarithmic score is now defined as the conditional predictive distribution given $\pmb{y}_{i}^{-j}$, i.e. $-\log(p_i(y_j|\pmb{y}_{i}^{-j}))$ \citep[see][for more details]{Cowell2007}.
The marginal and conditional node monitors test the respective probability distributions.

\paragraph{Parent-child monitors} 
After identifying problematic nodes, the parent-child monitor can be used to pinpoint the configurations of the parent values which might be associated with the misspecification. Consider a variable $Y_j$ and a specific value of its parents $y_{\pi_j}\in\mathbb{Y}_{\pi_j}$. Let $\pmb{y}_{\pi_j}=(\pmb{y}_{1}^{\pi_j},\dots,\pmb{y}_{m'}^{\pi_j})$  be the subvector of $\pmb{y}$ where only observations with $Y_{\pi_j}=y_{\pi_j}$ are retained. Consider the predictive distribution $p_i(y_{ij}^{\pi_j}|y_{\pi_j})$ after the first $i-1$-th observations of $\pmb{y}_{\pi_j}$ have been processed conditional on $y_{\pi_j}$ and define similarly to the node monitors
\begin{align*}
    E_{ij}^{\pi_j}&= - \sum_{y_j\in\mathbb{Y}_j}p_i(y_j|y_{\pi_j})\log(p_i(y_j|y_{\pi_j})), \\ V_{ij}^{\pi_j}&=\sum_{y_j\in\mathbb{Y}_j}p_i(y_j|y_{\pi_j})\log(p_i(y_j|y_{\pi_j}))^2 - (E_{ij}^{\pi_j})^2
\end{align*}
The sequential parent-child monitor for the vertex $Y_j$ and parent configuration $y_{\pi_j}$ is defined as
\[
Z_{ij}^{\pi_j}=\frac{-\sum_{k=1}^i\log(p_k(y_{kj}^{\pi_j}|y_{\pi_j}))-\sum_{k=1}^iE_{kj}^{\pi_j}}{\sqrt{\sum_{k=1}^iV_{kj}^{\pi_j}}}
\]
The interpretation of the parent-child monitor is the same as for the node monitors, where values above 1.96 should
be viewed with suspicion.
If the parent-child monitor indicates a poor fit for a particular set of parent values, this may indicate that a context-specific adaptation of the model would be appropriate (see the Discussion for more details).

\paragraph{Influential observations}
The influence of the $i$-th observation of $\bm{y}$ is defined as
\[
|\log(p_G(\pmb{y}))-\log(p_G(\pmb{y}_{-i}))|
\]
High values of influence denote observations that highly contribute to the likelihood of the model and therefore had a bigger impact during the selection of the DAG model.

\subsection{Sensitivity analysis in Bayesian networks}
Whilst the previous diagnostics are specifically designed for BNs learned for data, the next set of functions we illustrate apply to either data-learned or expert-elicited. No matter how the BN was derived, suppose we have a DAG $G$ and a probability distribution $p_G$ for $G$. We now review diagnostics which measure the impact of the conditional distributions $p(y_i|y_{\Pi_i})$ on output probabilities of interest, which are usually referred to as \emph{sensitivity analysis} \citep{Chan2002}.

Let $O,E\subset[n]$ be the index of the output and evidence variables, respectively. By output variable we mean a variable we may be interested in deriving its probability distribution, possibly conditional on a specific value of $y_E$ that we may observe. We are therefore interested in $p_G(y_O|y_E)$ and we want to study how this probability varies in terms of the conditional probabilities $p(y_i|y_{{\Pi}_i})$ that define the model. More specifically, the probability $p(y_O|y_E)$ seen as a function of $p(y_i|y_{{\Pi}_i})$ is called a \emph{sensitivity function} \citep[][]{Coupe2002}. These are used to assess if changes in the probabilities of the BN have a big impact on output probabilities of interest.

A similar task often performed in sensitivity analysis is the following. Suppose a probability of interest $p_G(y_O|y_E)$, although being computed from the conditional probabilities $p(y_i|y_{{\Pi}_i})$ defining $p_G$, takes a value which does not seem to be appropriate to the modeler. Further suppose the modeler has some idea of what value this probability should take, say a number $a$. The question then is the following: which changes in the conditional probabilities $p(y_i|y_{\Pi_i})$ would make $p_G(y_O|y_E)=a$? The issue was first addressed by \citet{Chan2002} and \texttt{bnmonitor} provides an implementation to answer this question.

Last, assume changes in the conditional distributions $p(y_i|y_{\Pi_i})$ have been identified, perhaps in order to apply a constraint as discussed in the previous paragraph. Denote the new probability distribution with such changes $p_G'$. It is then useful to assess what is the impact of these local changes of the conditional distributions on the overall BN distribution $p_G$. In other words, how far apart are $p_G$ and $p_G'$ from each other? There are various ways to measure such dissimilarity, including the Kullback-Leibler divergence and the Jeffreys distance \citep{Kullback1951}. A distance specifically designed for this task was introduce by \citet{Chan2005} and henceforth called \emph{CD distance}. The CD distance between two probability distributions $p_G$ and $p_G'$ is defined as
\[
\textnormal{CD}(p_G,p_G')=\log\max_{y\in\mathbb{Y}}\left(\frac{p_G(y)}{p_G'(y)}\right) - \log\min_{y\in\mathbb{Y}}\left(\frac{p_G(y)}{p_G'(y)}\right).
\]
All above mentioned measures are implemented in \texttt{bnmonitor} to quantify the overall effect of changes in the conditional probabilities of the BN.

\section{An applied analysis with \texttt{bnmonitor}}
We next illustrate the capabilities of the \texttt{bnmonitor} R package as well as the required syntax by carrying out an extensive model validation for a BN learned using the \texttt{diabetes} dataset bundled in the package. 

\subsection{Data description}
The \texttt{diabetes} dataset is a discretized versions of the famous Pima Indian Diabetes dataset from the UCI machine learning repository reporting medical reports of Pima-Indian women of at least 21 years of age\footnote{We chose this dataset because it best showcases the function of our monitors. However, we acknowledge that this data is used here without the consent of or compensation for the original Akimel O'odham participants.}. Observations with missing values were dropped and continuous variables were discretized into binary ones using the equal quantile method \citep{Nojavan2017}. The resulting \texttt{diabetes} dataset has 392 observations and the following nine variables.

\begin{itemize}
    \item \texttt{PREG}: number of times pregnant (low/high);
    \item \texttt{GLUC}: plasma glucose concentration (low/high);
    \item \texttt{PRES}: diastolic blood pressure (low/high);
    \item \texttt{TRIC}: triceps skin fold thickness (low/high);
    \item \texttt{INS}: 2-hour serum insulin (low/high);
    \item \texttt{MASS}: body mass index (low/high);
    \item \texttt{PED}: diabetes pedigree function (low/high);
    \item \texttt{AGE}: age (low/high);
    \item \texttt{DIAB}: test for diabetes (neg/pos).
\end{itemize}

The \texttt{DIAB} variable is an indicator (pos) for a positive test for diabetes between 1 and 5 years from
the examination determining the other variables, or (neg) a negative test for diabetes 5 or more years
later. 

\subsection{Learning a BN}
To start the analysis, we load all required packages as well as the \texttt{diabetes} dataset.
\begin{lstlisting}
library("bnmonitor")
library("bnlearn")
library("qgraph")
library("gRain")
data(diabetes)
\end{lstlisting}

The DAG of a BN is learned for the \texttt{diabetes} dataset using the hill-climbing algorithm implemented in the \texttt{hc} function of \texttt{bnlearn}. The visualization of the DAG is obtained via the \texttt{qgraph} function.

\begin{lstlisting}
dag <- hc(diabetes)
qgraph(dag)
\end{lstlisting}

\begin{figure}
    \centering
    \includegraphics[scale=0.9]{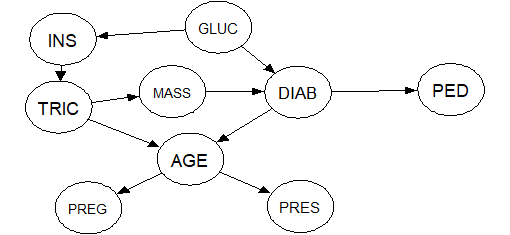}
    \caption{Learned BN with the \texttt{hc} function from \texttt{bnlearn} for the \texttt{diabetes} dataset.}
    \label{fig:my_label}
\end{figure}

The output is reported in Figure \ref{fig:my_label}. The DAG suggests that, for example, the outcome of the diabetes test is independent of the serum insulin and the triceps thickness given the body mass index and the glucose concentration. Similarly, the result of the pedigree function is independent of all other variables given the outcome of the diabetes test. The DAG further implies that \texttt{AGE} is independent of \texttt{PED}, \texttt{MASS}, \texttt{INS} e \texttt{GLUC} given \texttt{TRIC} and \texttt{DIAB}.

\subsection{Using robustness monitors}
We start the validation of the learned BN by computing the 
contributions of each individual node to the overall log-likelihood of the model via the function \texttt{global\_monitor}.

\begin{lstlisting}
global_monitor(dag = dag, df = diabetes)
\end{lstlisting}

\begin{table}
\centering
    \begin{tabular}{rr}
    \toprule
    Vertex     & Score \\
    \midrule
     PREG & 236.2658\\
   GLUC & 274.3482\\
  PRES & 250.0871\\
   TRIC &267.1841\\
   INS &219.8782\\
      MASS &231.8470\\
  PED &272.6041\\
       AGE &246.5046\\
  DIAB & 214.0108\\
  \bottomrule
    \end{tabular}
\caption{Global monitor for the diabetes network. \label{table:global}} 
\end{table}

The output is shown in Table \ref{table:global}. This can be used as the starting point for checking robustness, and in particular to compare models in the case there are competing ones. Although the node GLUC has a high contribution to the likelihood, this is a root of the BN, which is often not relevant for checking robustness \citep[see e.g.][]{Cowell2007}. Furthermore, we see the the pedigree function has a high contribution to the likelihood and for this reason we will investigate this node further, together with the outcome of the diabetes test.  

We start by computing the marginal and conditional node monitors for PED and DIAB using the functions \texttt{seq\_marg\_monitor} and \texttt{seq\_cond\_monitor}.

\begin{lstlisting}
plot(seq_marg_monitor(dag, diabetes, "DIAB"))
plot(seq_marg_monitor(dag, diabetes, "PED"))
plot(seq_cond_monitor(dag, diabetes, "DIAB"))
plot(seq_cond_monitor(dag, diabetes, "PED"))
\end{lstlisting}

\begin{figure}
    \centering
    \includegraphics[scale=0.30]{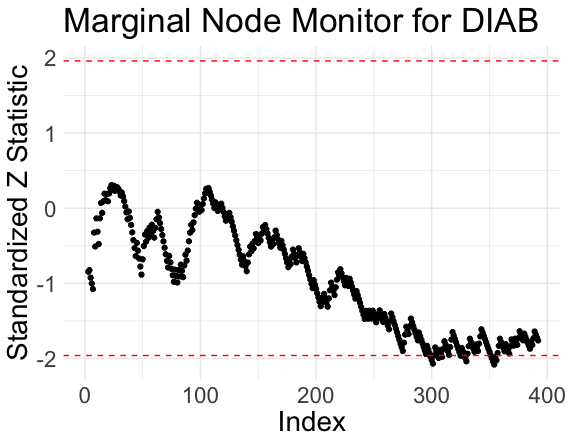}\; \includegraphics[scale=0.30]{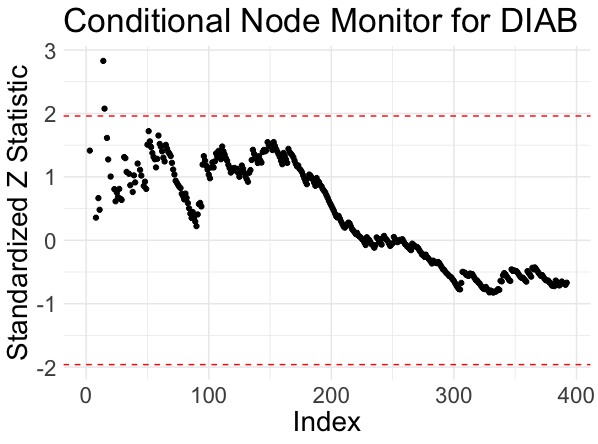}\\
    \includegraphics[scale=0.30]{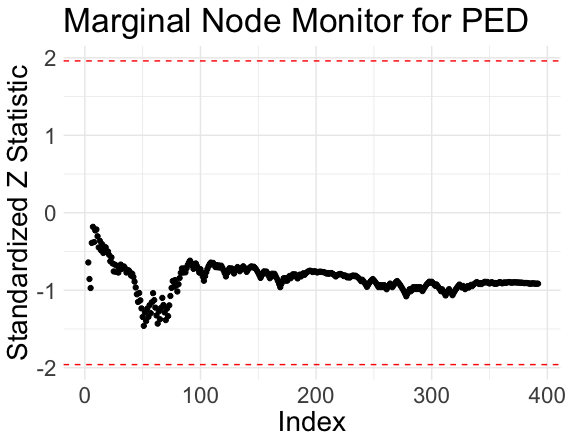}\;
    \includegraphics[scale=0.30]{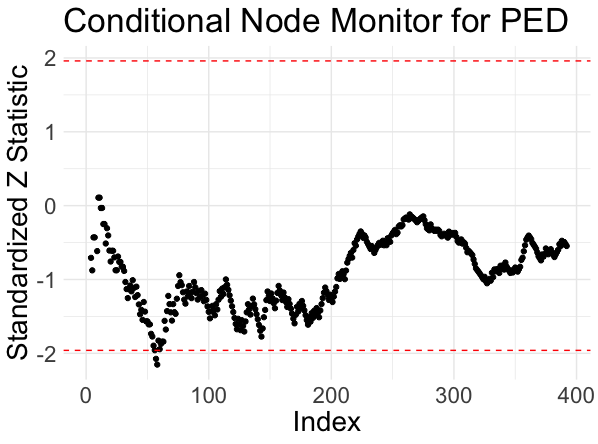}
    \caption{Node monitors for the vertices diabetes and pedigree. The red dashed line indicates $|Z_i|=1.96$}
    \label{fig:node}
\end{figure}

The output in Figure \ref{fig:node} shows that the diabetes pedigree function is robust and a good fit for the data, since the monitors are almost always the 1.96 confidence bands.  Conversely, the marginal monitor for the test for diabetes indicates that forecasts from $i \in (300,396)$ may be suspect.

Unpicking the reason for this shift in the accuracy of the forecasts may be due to a latent variable in the data generating process. There are two possible ways to investigate further this lack of fit. One is to construct parent-child monitors. The other is to reorder the data  and rerun the node monitors: this could reveal an additional dependence obscuring the forecasts. Here we follow the first route.

\begin{figure}
     \centering
     \begin{subfigure}[b]{0.48\textwidth}
         \centering
         \includegraphics[width=\textwidth]{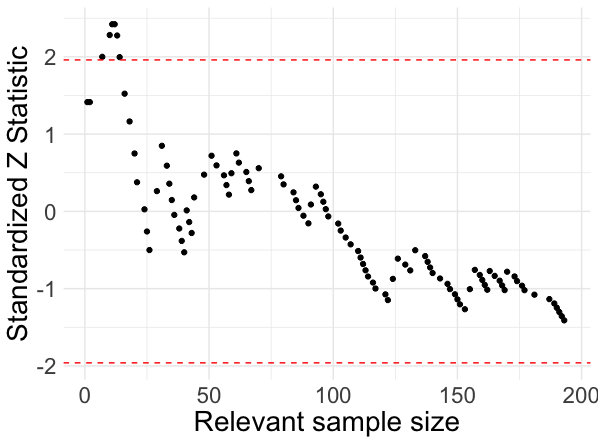}
         \caption{(GLUC, MASS) = (low, low)}
         \label{fig:ll}
     \end{subfigure}
     \hfill
     \begin{subfigure}[b]{0.48\textwidth}
         \centering
         \includegraphics[width=\textwidth]{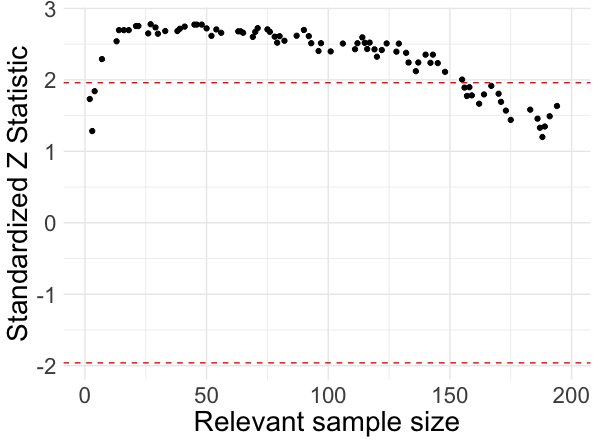}
         \caption{(GLUC, MASS) = (low, high)}
         \label{fig:lh}
     \end{subfigure}
\\
       \centering
     \begin{subfigure}[b]{0.48\textwidth}
         \centering
         \includegraphics[width=\textwidth]{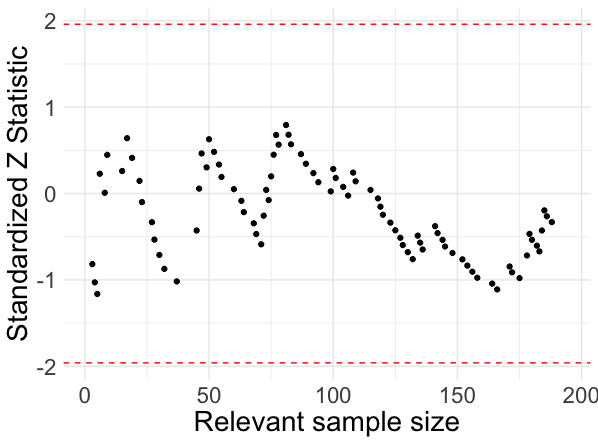}
         \caption{(GLUC, MASS) = (high, low)}
         \label{fig:hl}
     \end{subfigure}
     \hfill
     \begin{subfigure}[b]{0.48\textwidth}
         \centering
         \includegraphics[width=\textwidth]{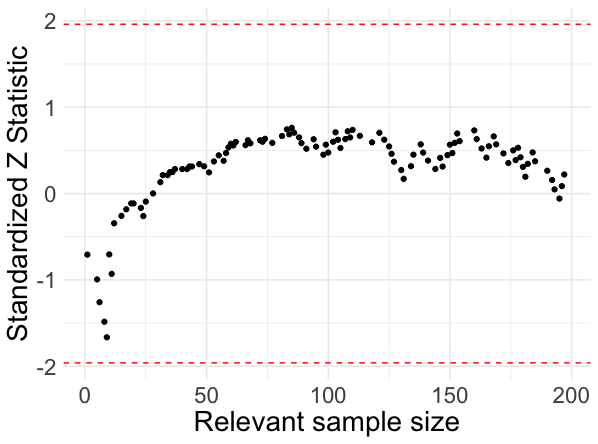}
         \caption{(GLUC, MASS) = (high, high)}
         \label{fig:hh}
     \end{subfigure}
        \caption{Parent-child monitors for the diabetes vertex.}
        \label{fig:pach1}
\end{figure}

Parent-child monitors examine the subset of the data which has a specific value of the parents of a node and examines the forecasts that flow from that subset.
The parents of diabetes are plasma glucose concentration and body mass index.  For instance, to construct the parent-child monitor when the parents take the low value we use the code:

\begin{lstlisting}
plot(seq_pa_ch_monitor(dag, diabetes, "DIAB",
 pa.names = c("GLUC","MASS"), pa.val = c("low","low")))
\end{lstlisting}

Figure \ref{fig:pach1} shows the monitors for all parent configurations of the node DIAB. The model struggles to forecast the outcome of the diabetes test for participants who have low plasma glucose concentration and high BMI since Figure \ref{fig:ll} shows plots with a $Z_i > 1.96$. Conversely, the forecasts for the other parents' configurations are accurate. In some cases, it may be appropriate to create a context-specific BN for the problematic parent values (see the Discussion below for more details).

\begin{figure}
     \centering
     \begin{subfigure}[b]{0.48\textwidth}
         \centering
         \includegraphics[width=\textwidth]{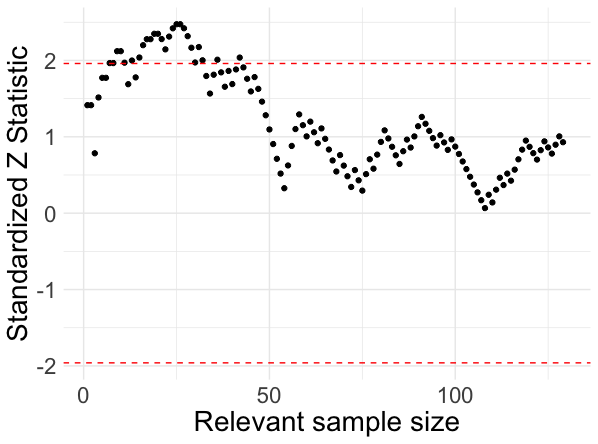}
         \caption{DIAB = pos}
         \label{fig:p}
     \end{subfigure}
     \hfill
     \begin{subfigure}[b]{0.48\textwidth}
         \centering
         \includegraphics[width=\textwidth]{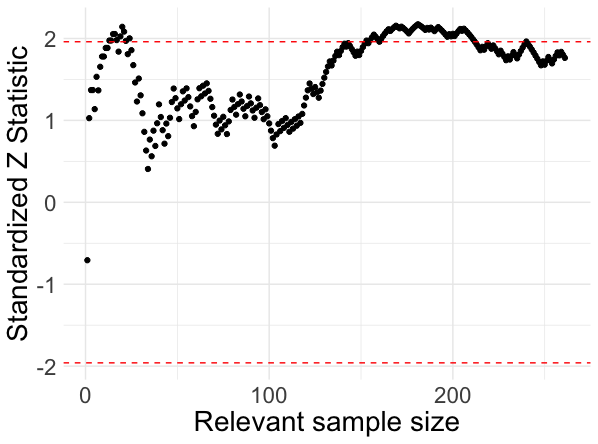}
         \caption{DIAB = neg}
         \label{fig:n}
     \end{subfigure}
      \caption{Parent-child monitors for the pedgree vertex.}
        \label{fig:pach2}
        \end{figure}

Figure \ref{fig:pach2} shows the parent-child monitors for the pedigree function node.  
Figure \ref{fig:p} shows problematic forecasts towards the beginning of the relevant sample size. 
This could be due to the 'burn-in' from learning the model. 
Conversely, the forecasts at the end of the relevant sample size in Figure \ref{fig:n} are close to the confidence bands and therefore merit further inquiry.

As a last check of the model fit to data, we compute the influence of the observations via the \texttt{influential\_obs} function.
The most influential observations can be thought of as the most unusual.

\begin{lstlisting}
influence <- influential_obs(dag, diabetes)
plot(influence)
\end{lstlisting}

\begin{figure}
    \centering
    \includegraphics[scale = 0.8]{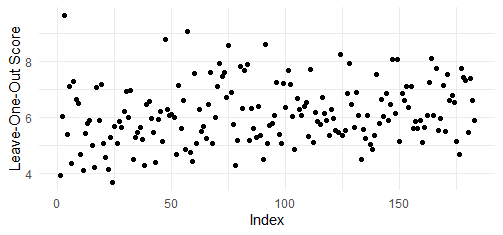}
    \caption{Influence score of the observations in the diabetes dataset.}
    \label{fig:influence}
\end{figure}

Figure \ref{fig:influence} shows that there are quite a few observations that highly contribute to the likelihood of the model (influence over 8.5). The actual observations with high influence can be derived as follows. 

\begin{lstlisting}
subset(unique(influence),score > 8.5)
\end{lstlisting}

The output is shown in Table \ref{table:infl}. It can be seen that the observations which highest influence are all for low levels of glucose and a positive diabetes test.

\begin{table}
\centering
\scalebox{0.9}{
\begin{tabular}{cccccccccc}
\toprule
    PREG & GLUC &PRES& TRIC & INS &MASS & PED & AGE &DIAB &   score \\
    \midrule
  high & low & low &high & low & low & low & low&  pos& 9.652\\
  low & low & high & low & low & high & low & low & pos & 8.811\\
 low & low  & high  & low  & high & high & low& high & pos &9.070\\
high & low & low & low & low & low &high & low & pos &8.581\\
 high & low & low & low&high & low &high& high&  pos& 8.609\\
 \bottomrule
\end{tabular}}
\caption{Most influential observations in the diabetes dataset. \label{table:infl}}
\end{table}

\subsection{Checking the learned probabilities}

The previous analysis demonstrated that overall a BN provides a good fit for the \texttt{diabetes} dataset. Next let's check the implications of the model. First, we need to create a \texttt{bn.fit} object that we call \texttt{bn}.

\begin{lstlisting}
bn <- bn.fit(dag, diabetes)
\end{lstlisting}

The vertex of most interest is \texttt{DIAB} reporting the result of a diabetes test (either positive or negative). As an illustration, we first investigate how the probability of a positive test depends on the variable \texttt{GLUC} = high. This can be done via the function \texttt{sensitivity} whose output can be plotted using the \texttt{plot} method.

\begin{lstlisting}
sens_dc <- sensitivity(bn, interest_node = "DIAB",
    interest_node_value = "pos", node = "GLUC", 
    value_node = "high", value_parents = NULL,
    new_value = "all")
plot(sens_dc)
\end{lstlisting}

\begin{figure}
     \centering
     \begin{subfigure}[b]{0.48\textwidth}
         \centering
         \includegraphics[width=\textwidth]{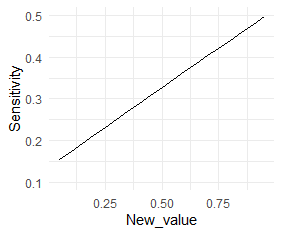}
         \caption{\texttt{sens\_dc}}
         \label{fig:sens1}
     \end{subfigure}
     \hfill
     \begin{subfigure}[b]{0.48\textwidth}
         \centering
         \includegraphics[width=\textwidth]{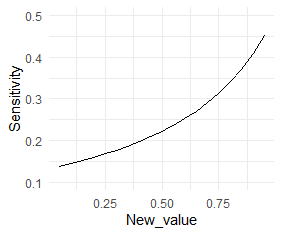}
         \caption{\texttt{sens\_dic}}
         \label{fig:sens2}
     \end{subfigure}
      \caption{Plots of the sensitivity functions for the \texttt{diabetes} dataset.}
        \label{fig:sens}
        \end{figure}

The output is reported in Figure \ref{fig:sens1}. The plot shows that has the probability of having a high level of glucose increases, then also the probability of a positive test increases. Notice that in this case, since \texttt{GLUC} is a root vertex, we set  the input \texttt{value\_parents} equal to \texttt{NULL}.

In the previous example the probability of interest was the marginal probability of a positive test. Similarly, we can assess how generic conditional probabilities are affected by changes in the model. As an illustration, let's consider the conditional probability of a positive test given a low level of insulin and investigate how this varies when the probability of a high level of glucose changes. This can be done similarly to the previous code, but now we have to fix the \texttt{evidence\_nodes} and \texttt{evidence\_states} inputs.

\begin{lstlisting}
send_dic <-  sensitivity(bn, interest_node = "DIAB", 
    interest_node_value =  "pos",
    evidence_nodes = "INS", evidence_states = "low",
    node = "GLUC", value_node = "high",
    value_parents = NULL, new_value = "all")
plot(sens_dic)
\end{lstlisting}

The output is reported in Figure \ref{fig:sens2} and again the output (conditional) probability of a positive test increases when the probability of a high level of glucose increases. Notice that in this case the increase is non-linear as expected from the results of \citet{Coupe2002}.

We now might be interested in knowing how much changes in the probability of high glucose affect the overall probability distribution of the BN. We compute the CD distance using the \texttt{CD} function and the associated \texttt{plot} method. 

\begin{figure}
     \centering
     \begin{subfigure}[b]{0.48\textwidth}
         \centering
         \includegraphics[width=\textwidth]{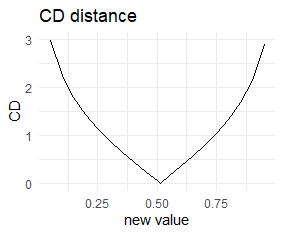}
         \caption{\texttt{cd\_g}}
         \label{fig:CD1}
     \end{subfigure}
     \hfill
     \begin{subfigure}[b]{0.48\textwidth}
         \centering
         \includegraphics[width=\textwidth]{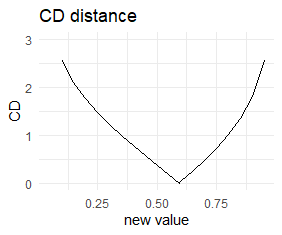}
         \caption{\texttt{cd\_d}}
         \label{fig:CD2}
     \end{subfigure}
      \caption{Plots of the CD distances for the \texttt{diabetes} dataset.}
        \label{fig:CD}
        \end{figure}

\begin{lstlisting}
cd_g <- CD(bn, node = "GLUC", value_node = "high",
    value_parents = NULL, new_value = "all")
plot(cd_g)
\end{lstlisting}
The output is given in Figure \ref{fig:CD1} and one can see that the original value of this probability was around 0.5 since the CD is zero at that point.

As an illustration, let's  also consider the CD distance when the conditional probability of a positive diabetes test given a high body mass index and a high glucose is varied.
\begin{lstlisting}
cd_d <- CD(bn, node = "DIAB", value_node = "pos", 
    value_parents = c("high","high"), new_value = "all")
plot(cd_d) 
\end{lstlisting}
The plot is reported in Figure \ref{fig:CD2} and we can notice that overall the CD distance is smaller for changes of this probability compared to the one in Figure \ref{fig:CD1}.

Last, we may check if probabilities implied by the BN are reasonable. Let's consider as an example the conditional probability of a positive diabetes test given a high pressure level. We can compute this probability using the \texttt{querygrain} function from the \texttt{gRain} package.

\begin{lstlisting}
querygrain(as.grain(bn), nodes = c("DIAB","PRES"), 
    type = "conditional")
\end{lstlisting}
The probability is estimated as 0.38, which appears to be lower than plausibly expected. Suppose we believe this probability should on the other hand be at least 0.4. The function \texttt{sensquery} outputs the probability changes which would make this probability at least 0.4.

\begin{lstlisting}
sensquery(bn, interest_node = "DIAB",
    interest_node_value = "pos", new_value = 0.4, 
    evidence_nodes = "PRES", evidence_states = "high")
\end{lstlisting}

\begin{table}[]
    \centering
    \scalebox{0.84}{
    \begin{tabular}{rrrrrr}
    \toprule
  Node Value & Node & Value parents & Original value  & Suggested change & CD distance \\
  \midrule
 GLUC    &    low &      &            0.4872445   &     0.44091  & 0.18642 \\
 PRES  &       low     &      low    & 0.608466   &     0.69930  & 0.40310\\
DIAB    &    pos   &    low,low   &  0.056604     &   0.12506  & 0.86801\\
MASS   &    high     &      low    & 0.260638    &    0.48177  & 0.96970\\
  AGE  &      low  &     low,neg    & 0.703448     &   0.92669  & 1.67311\\
  \bottomrule
    \end{tabular}
    }
    \caption{Possible probability changes output of \texttt{sensquery}}
    \label{tab:query}
\end{table}
The output is reported in Table \ref{tab:query}. There are five possible changes in the conditional probabilities of the model meeting this constraint. The function further reports the CD distance associated to the changes, which ranges from 0.19 up to 1.67.

\section{Discussion}
We demonstrated the usage of the \texttt{bnmonitor} R package using a real-world dataset and the insights that modelers can get by carrying out a model validation analysis. We highlighted both the plotting capabilities of \texttt{bnmonitor} which take advantage of the R environment and seamless integration with the famous \texttt{bnlearn} package.

As already mentioned, \texttt{bnmonitor} can also deal with continuous random variables and provides a wide array of functions to validate Gaussian BNs. Furthermore, since in our application all variables where binary, when we varied a probability by default the other is set as one minus the new value. However, in the generic categorical case there are various methods to adjust probabilities  \citep{Renooij2014} which are available and implemented in \texttt{bnmonitor}.

One issue we did not address in this paper is in providing alternatives to the modeller in the case the BN does not provide a good representation of the data. One possibility is to consider a more flexible model generalizing BNs, namely the class of staged tree models \citep{Smith2008}. There is now an R package called \texttt{stagedtrees} which implements structural learning algorithm for this model class \citep{Carli2020}. We are currently developing an additional R package implementing similar diagnostics which have been developed in \citet{Leonelli2017} and \citet{Wilkerson2019}.



\bibliographystyle{elsarticle-harv} 
\bibliography{Bib}





\appendix

\section{Predictive density}
\label{sec:app}
Recall that we consider a BN over a random vector $Y=(Y_1,\dots,Y_n)$, whose sample space is $\mathbb{Y}=\times_{i=1}^n\mathbb{Y}_i$ and $\mathbb{Y}_i$ is the sample space of $Y_i$. 
We denote a generic probability from the BN as $\theta_{ijk}=p(k|j)$ where $k\in\mathbb{Y}_i$ and $j\in \mathbb{Y}_{\Pi_i}$. 

Suppose we observed a dataset $\pmb{y}$ and let $N_{ijk}$ the number of observations in $\pmb{y}$ such that $Y_i=k$ and $Y_{\Pi_i}=j$. Then the likelihood of the BN is 
\[
L(\theta|\pmb{y})= c\prod_{i\in [n]}\prod_{j\in \mathbb{Y}_{\Pi_i}}\prod_{k\in \mathbb{Y}_i}\theta_{ijk}^{N_{ijk}},
\]
where $c$ is a normalizing constant and $\theta$ is the vector of all probability parameters $\theta_{ijk}$.

The monitors we discussed in Section \ref{sec:rob} are all within the Bayesian inferential framework which requires the definition of a prior distribution for $\theta$. We independently assign to each $\theta_{ij}=(\theta_{ijk})_{k\in\mathbb{Y}_i}$ a Dirichlet prior $\mathcal{D}(\alpha_{ijk})_{k\in\mathbb{Y}_i}$ where $\alpha_{ijk}>0$. The overall prior distribution for $\theta$ can then be written as
\[
p(\bm{\theta})= \prod_{i\in [n]} \prod_{j\in\mathbb{Y}_{\Pi_j}} \frac{\Gamma\left(\sum_{k\in\mathbb{Y}_i} \alpha_{ijk}\right)}{\prod_{k\in\mathbb{Y}_i} \Gamma(\alpha_{ijk})} \prod_{k\in \mathbb{Y}_i} \theta_{ijk}^{\alpha_{ijk}-1},
\]
where $\Gamma(\cdot)$ denotes the Gamma function.

Results first appeared in \citet{Heckerman1995} then guarantee that the posterior distribution $p(\theta|\pmb{y})$ can be written as
\[
p(\theta | \pmb{y}) = c \prod_{i\in [n]} \prod_{j\in\mathbb{Y}_{\Pi_i}} \frac{\Gamma\left(\sum_{k\in\mathbb{Y}_i} \alpha_{ijk}+N_{ijk}\right)}{\prod_{k\in\mathbb{Y}_i} \Gamma(\alpha_{ijk}+N_{ijk})} \prod_{k\in \mathbb{Y}_i} \theta_{ijk}^{\alpha_{ijk}+N_{ijk}-1},
\]
where $c$ is again a normalizing constant. The above expression can be seen as the product of independent Dirichlet distributions for $\theta_{ij}$ $\mathcal{D}(\alpha_{ijk}+N_{ijk})_{k\in\mathbb{Y}_i}$. 

Lastly, the predictive distribution $p(\bm{y})$ can be derived from the posterior distribution $p(\theta|\pmb{y})$ and is equal to
\[
p(\pmb{y})=\prod_{i\in[n]}\prod_{j\in\mathbb{Y}_{\Pi_{j}}}\frac{N_{ij}!}{\sum_{k\in\mathbb{Y}_i}(N_{ijk}!)}\frac{\Gamma(\alpha_{ij})}{\prod_{k\in\mathbb{Y}_j}\Gamma(\alpha_{ijk})}\frac{\prod_{k\in\mathbb{Y}_i\Gamma(\alpha_{ijk}+N_{ijk})}}{\Gamma(\alpha_{ij}+N_{ij})},
\]
where $\alpha_{ij}=\sum_{k\in\mathbb{Y}_i}\alpha_{ijk}$. The above expression underlies the computation of all monitors of Section \ref{sec:rob} and requires a user's choice for the values of $\alpha_{ijk}$. By default, these are given the value $|\mathbb{Y}_i|$, i.e. the number of elements in the sample space of $Y_i$.

\end{document}